\documentclass[sigconf,nonacm]{acmart}

\usepackage{graphicx,xcolor,footmisc,times}
\usepackage[export]{adjustbox}

\usepackage{subfig}
\usepackage{multirow}
\usepackage{float}

\begin{document}
\copyrightyear{2022}
\acmYear{2022}
\setcopyright{acmcopyright}\acmConference[WWW '22 Companion]{Companion Proceedings of the Web Conference 2022}{April 25--29, 2022}{Virtual Event, Lyon, France}
\acmBooktitle{Companion Proceedings of the Web Conference 2022 (WWW '22 Companion), April 25--29, 2022, Virtual Event, Lyon, France}
\acmPrice{15.00}
\acmDOI{10.1145/3487553.3524642}
\acmISBN{978-1-4503-9130-6/22/04}


\title{TweetBoost: Influence of Social Media on NFT Valuation}

\author{Arnav Kapoor$^*$}
\author{Dipanwita Guhathakurta$^*$}
\affiliation{
  \institution{IIIT Hyderabad}
  \country{}
}

\email{arnav.kapoor@research.iiit.ac.in }
\email{dipanwita.g@research.iiit.ac.in}

\thanks{$^*$ denotes equal contribution}

\author{Mehul Mathur$^*$}
\author{Rupanshu Yadav$^*$}

\affiliation{
  \institution{IIIT Hyderabad}
  \country{}
}
\affiliation{
  \institution{IIIT Delhi}
  \country{}
}

\email{mehul.mathur@students.iiit.ac.in}
\email{rupanshu19475@iiitd.ac.in}

\author{Manish Gupta}
\author{Ponnurangam Kumaraguru}

\affiliation{
  \institution{IIIT Hyderabad}
  \country{}
}
\email{manish.gupta@iiit.ac.in}
\email{pk.guru@iiit.ac.in}

\renewcommand{\shortauthors}{Kapoor$^*$, Guhathakurta$^*$, Mathur$^*$, Yadav$^*$, Gupta and Kumaraguru}

\begin{abstract}

NFT or Non-Fungible Token is a token that certifies a digital asset to be unique. A wide range of assets including, digital art, music, tweets, memes, are being sold as NFTs. NFT-related content has been widely shared on social media sites such as Twitter. We aim to understand the dominant factors that influence NFT asset valuation. Towards this objective, we create a first-of-its-kind dataset linking Twitter and OpenSea (the largest NFT marketplace) to capture social media profiles and linked NFT assets. Our dataset contains 245,159 tweets posted by 17,155 unique users, directly linking 62,997 NFT assets on OpenSea worth 19 Million USD. We have made the dataset publicly available.\footnote{\url{https://tinyurl.com/NFTValuation}\label{datafootnote}}

We analyze the growth of NFTs, characterize the Twitter users promoting NFT assets, and gauge the impact of Twitter features on the virality of an NFT. Further, we investigate the effectiveness of different social media and NFT platform features by experimenting with multiple machine learning and deep learning models to predict an asset's value. We model the problem as a binary classification as well as an ordinal classification task. Our results show that social media features improve the ordinal classification accuracy by 6\% over baseline models that use only NFT platform features. Among social media features, count of user membership lists, number of likes and replies are important features. On the other hand, OpenSea features like offer entered, bids withdrawn, bid entered and is presale turn out to be important predictors.

\end{abstract}




\maketitle

\section{Introduction}
Blockchain has emerged as a core disruptive technology that has transformed the financial ecosystem. The origin of Blockchain can be traced back to a 2008 whitepaper~\cite{nakamoto2008bitcoin} published under the pseudonym Satoshi Nakamoto, who introduced blockchain in the context of the most popular crypto-currency, Bitcoin. Bitcoin uses Blockchain technology to develop the publicly distributed ledger used to record the transactions on its network.
The growing interest in Blockchain technology especially its use in the financial domain from both retail and institutional investors\footnote{\url{https://www.forbes.com/sites/lawrencewintermeyer/2021/08/12/institutional-money-is-pouring-into-the-crypto-market-and-its-only-going-to-grow/}} has led to several new products emerging in the crypto-sphere to find the `next big thing'. 

One such emerging Blockchain product that has captured large public attention is Non-Fungible Tokens or NFTs. An NFT is a token that certifies a digital asset to be unique. NFTs use blockchain to store anything that can be converted into digital files, for example, images, music, and videos. Blockchain technology enables the association of proof of ownership to the digital asset. NFTs have grown exponentially in 2021; this phenomenal growth has led to traditional auction houses being receptive to digital art NFTs. On March 11, 2021, \textit{`Everydays: The First 5000 Days'}, an NFT artwork of the prominent digital artist Beeple, sold at Christie's for over \$69 million.\footnote{\url{https://www.theverge.com/2021/3/11/22325054/beeple-christies-nft-sale-cost-everydays-69-million}} Jack Dorsey (CEO of Twitter) raised an astounding \$2.9 million for charity by auctioning his first tweet as an NFT.\footnote{\url{https://www.cnbc.com/2021/03/22/jack-dorsey-sells-his-first-tweet-ever-as-an-nft-for-over-2point9-million.html}} All these sales demonstrate the adoption of NFTs by the mainstream community. 
The highly volatile NFT prices and sudden popularity led many people to create NFTs and in turn, promote and sell them for a profit. The most significant impact of NFTs has been on how they transformed the art world~\cite{van_haaften-schick_artists_2020,whitaker_art_2019}. NFTs have allowed artists to sell their art outside the gate-keeping systems and taste-hierarchies~\cite{van_haaften-schick_artists_2020}. 
NFT is a token representing digital assets stored on the blockchain with proof of ownership. Smart contracts \cite{ante_smart_2021} allow to transfer and retrieve information about an NFT through function calls. Some function calls such as `transfer' are restricted to the owner of the NFT. Since the underlying blockchain is decentralized, no one can change the state of the contract or the NFT without making these function calls, ensuring high levels of security.

Most NFTs are digital, meaning that consumers do not receive any physical items when they purchase them. In most circumstances, the NFT is only a proof of ownership, not of copyright; i.e., the owner does not have exclusive access to the content of NFTs. For example, `disaster girl’, a popular Internet meme, was sold as an NFT for \$495,000 even though the exact image in the NFT is freely available and distributed throughout the Internet.\footnote{\url{https://www.nytimes.com/2021/04/29/arts/disaster-girl-meme-nft.html}} The value of the NFT came from the fact that it was sold by the girl featured in the meme. The same meme sold as an NFT by other people on the marketplace did not receive any traction.\footnote{\url{https://bit.ly/3CLf3W8}}  Interestingly, most NFTs sold online can be downloaded and shared publicly for free. The value of an NFT is based on the perception of buyers which arises from the recognition of the creator and the overall marketing around the NFT itself. Further, unlike company stocks or crypto-currency exchanges, which are traded at regular intervals, NFTs have sporadic sales. 
The transaction history of an NFT spans varied durations and the owner changes in each sale. A single sale cannot account for the overall value of an NFT as the next buyer may be willing to pay much more or less than the previous sale amount depending on their current perception of the NFT. Hence, we use the average of all sales to assign a value to the NFT. The tremendous volatility of crypto-currencies, the lack of any tangible asset and the speculative marketspace make the asset valuation task for NFTs extremely challenging. Unlike traditional assets, NFT asset valuation cannot be modelled directly as a mathematical economic system, but rather as a social phenomenon involving marketing schemes and the recognition and popularity of the NFT. 

There are various marketplaces to buy and sell different categories of NFTs, such as OpenSea\footnote{\url{https://opensea.io}}, Rarible\footnote{\url{https://rarible.com}}, SuperRare\footnote{\url{https://superrare.com}}, NiftyGateway\footnote{\url{https://niftygateway.com}} and Foundation\footnote{\url{https://foundation.app}}.
OpenSea is the largest and most popular of all
such marketplaces, with over 300,000 users with \$3.4 Billion volume traded in August 2021 alone. Hence, we chose OpenSea to understand and model the NFT market.

\begin{figure}
    \centering
    \subfloat[Tweet announcing an NFT]{{\includegraphics[width=6.5cm,frame]{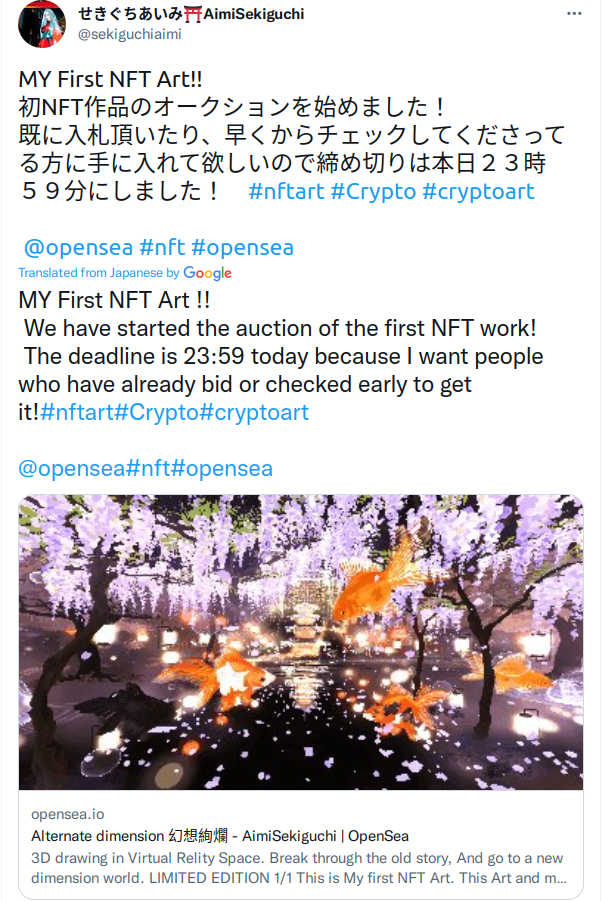}}}
    \qquad
    \subfloat[Corresponding OpenSea listing page]{{\includegraphics[width=6.5cm,height=7cm,frame]{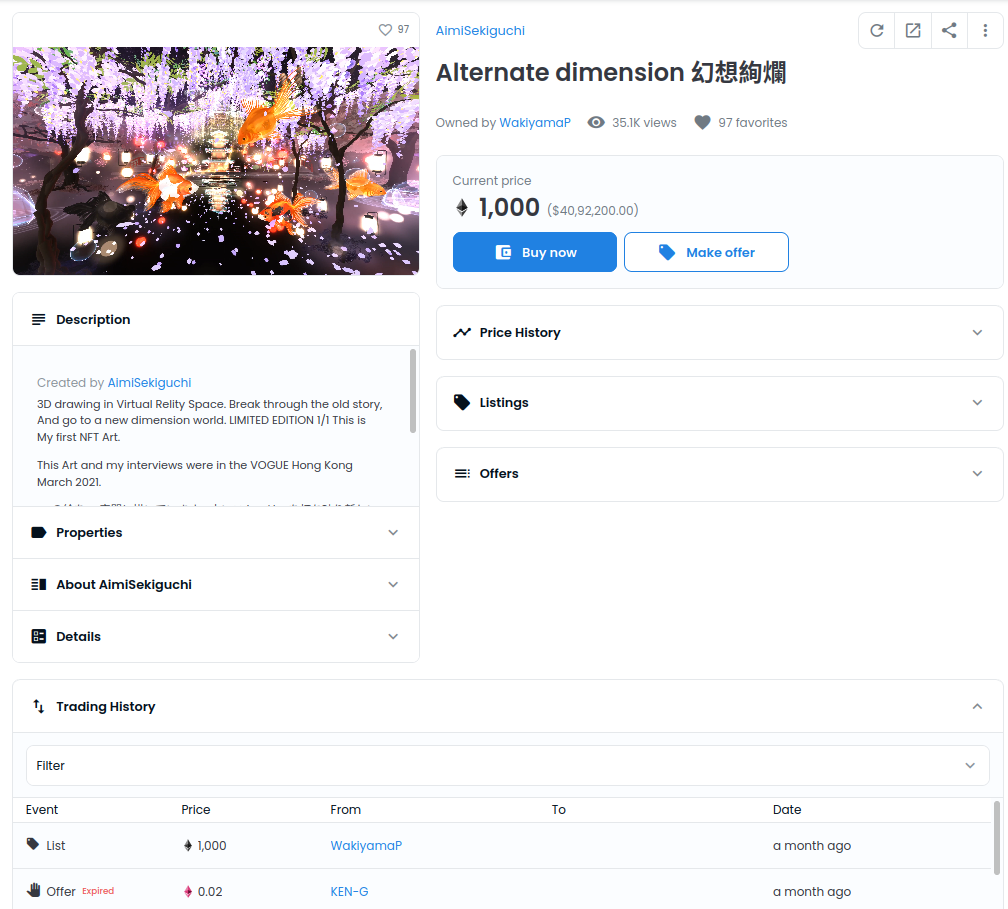}}}
    \caption{One of the most-liked tweets in our dataset from Japanese VR artist Aimi Sekiguchi and its corresponding page on OpenSea. This 14-second NFT video sold for over \$150,000  on March 24, 2021.}
    \label{fig:Sample_tweet-asset}
\end{figure}

To explore the branding and context around the NFT, we look at social media, particularly Twitter as a vehicle for building public perception and attracting potential buyers. Fig.~\ref{fig:Sample_tweet-asset}(a) shows a popular tweet and the corresponding NFT asset sold for over \$150,000  on OpenSea. Several instances of well-known personalities like Mark Cuban, Jack Dorsey, etc. 
on Twitter selling high-priced NFTs indicate that social media reach can play a role in influencing asset value. More than 70\% of the total traffic from social media on OpenSea is from Twitter.\footnote{\url{https://www.similarweb.com/website/opensea.io/\#social}} Thus, we focus on Twitter to understand the influence of social media on the NFT market. We aim to assess if an asset is overvalued or undervalued based on our current asset valuation framework.

In this paper, we aim to answer the following research questions:
 \begin{itemize}
    \item RQ1. What is the relationship between user activity on Twitter and price on OpenSea?

    \item RQ2. Can we predict NFT value using signals obtained from Twitter and OpenSea; and identify which features have the greatest impact on prediction?
    
 \end{itemize}
  
\begin{figure*}
\centering
\includegraphics[width=0.9\linewidth]{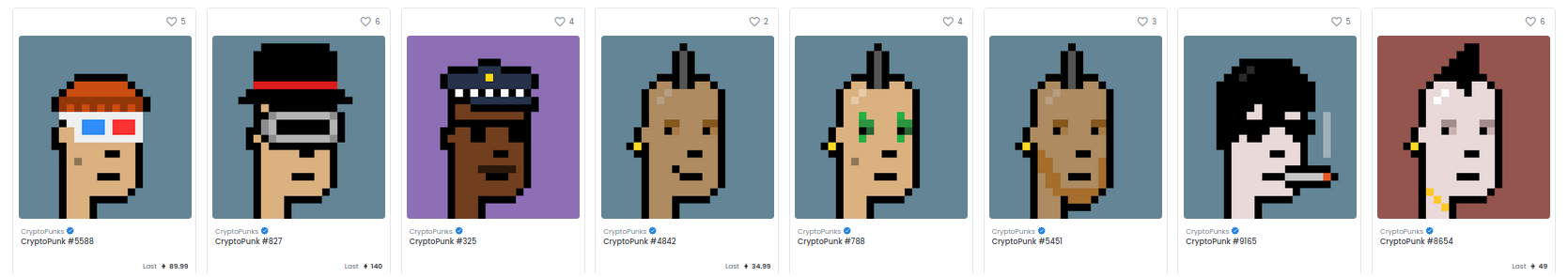}
\caption{CryptoPunks collection: A set of 10,000 assets showing similar traits.}
\label{fig: Cryptopunks}
\end{figure*}

We collect tweets announcing NFTs and follow the linked OpenSea URLs to extract NFT sales information and metadata from OpenSea. Fig.~\ref{fig:Sample_tweet-asset} shows an example of a tweet announcing an NFT and its corresponding OpenSea listing page. In addition, we crawl NFT images to analyse whether the image in itself exerts an influence on the price. Analysing growth and price trends on OpenSea against popularity metrics from Twitter reveals the possible significance of social media features on NFT value. We motivate average selling price as the metric to assign value to an NFT due to limited sales and highly volatile prices of NFT. We further develop and analyse predictive models using Twitter, OpenSea and NFT image features to assess their impact on asset value. Overall, we make the following contributions in this paper:

\begin{itemize}
    \item To the best of our knowledge, we create the first ever data set for NFTs from OpenSea and their corresponding tweets. We have made the dataset public\footref{datafootnote} in adherence to the FAIR (Findable, Accessible, Interoperable, Reusable) principles.

    \item We build ordinal classification models to predict NFT asset value using features from both OpenSea and Twitter, as well as the NFT image itself. Our best model comprising of an ensemble of Twitter and OpenSea features achieves an accuracy of 69.5\% in a 5-class classification setup. 
    
    \item We show that both Twitter  and OpenSea features influence the model output. In contrast the predictive power of image features is limited. This shows how branding and metadata (Twitter and OpenSea features) have a stronger effect on the value compared to the NFT product itself (Image features).

\end{itemize}

The remainder of the paper is organized as follows. 
In Section~\ref{sec:relatedWork}, we discuss related work on NFTs and asset valuation in the financial domain. We motivate and present our asset valuation problem setup in Section~\ref{sec:motivate_problem}. Section~\ref{sec:opensea_platform} briefly introduces the OpenSea platform and its features, as well as blockchain-specific terms used in our analysis. We discuss our data collection pipeline and analyse the impact of Twitter on the OpenSea marketplace from the temporal dimension and value perspective in Section~\ref{sec:data}. In Section~\ref{sec:Models}, we discuss multimodal models for NFT asset valuation. We present results using various models and feature subsets in Section~\ref{sec:results}. Finally, we conclude with limitations and future work in Section~\ref{sec:conclusion}.

\section{Related Work} 
\label{sec:relatedWork}

Since NFTs are a recent phenomena, there is limited research on NFT related data. Most of the previous work on NFTs is on analysis of relationship between blockchain, crypto-currency and NFTs. 
Recent work has analyzed the protocols, standards, desired properties, security, and challenges associated with NFTs~\cite{chevet_blockchain_2018,wang_non-fungible_2021}. Sudden rise of NFTs sparked an interest to find a correlation with more traditional crypto-assets. There has been a focus on understanding the correlation between the entire NFT market with Ethereum and Bitcoin~\cite{ante_non-fungible_2021} and also specific collections like Axie, CryptoPunks, and Decentraland~\cite{dowling_is_2021}. These sub-markets have millions of dollars traded every day~\cite{dowling_is_2021}, which has led to critical analysis of these individual markets. Other studies focus on fairness in NFT submarkets like CryptoKitties~\cite{sako_fairness_2021}.

To the best of our knowledge, there is limited work on NFT asset valuation, which is the focus of this work. Recently, Nadini et al.~\cite{nadini_mapping_2021} used simple machine learning algorithms to develop a predictive model using sale history and visual features, but ignore social media features. 
Social media features have helped predict the prices of assets in traditional markets~\cite{bollen_Twitter_2011,pineiro-chousa_influence_2017} like stock prices~\cite{nguyen_sentiment_2015}. 
Hence, in this work, we focus on understanding impact of OpenSea and social media features for this task. 

\section{The NFT Asset Valuation Problem}
\label{sec:motivate_problem}
NFT markets are highly illiquid in nature, which means sale prices are very volatile and irregular. Similar to physical art collections, the number of transactions per NFT is low as the buyers and sellers are a small niche of collectors. A traditional price prediction setting is not feasible or robust for each NFT as we do not have consistent periodic sales. In our dataset as well, a very small percentage (0.9\%) of the total assets had more than 5 sales. Price prediction models used for stock or cryptocurrencies use a large number of historical data points gathered at regular intervals which is not available for an illiquid market like NFT. 

We propose an asset valuation task instead of a price prediction task to overcome the challenge of an illiquid market.   
We define asset value as the average selling price of an asset over all its historic sales.
This compensates the large volatility in selling price, and provides a better indicator of asset valuation. Our objective is to provide a quantitative assessment of the NFT's value and identify whether the Twitter reach of the NFT influences the value. We thus divide the NFTs into multiple asset classes ( Very Low Value Asset, Low Value Asset, Medium Value Asset, High Value Asset, Very High Value Asset) 
based on the average sale price. For sake of brevity, we will refer to these classes as Class1, Class2, Class3, Class4 and Class5 respectively.
The maximum average price was found to be orders of magnitude more than the minimum, hence the asset classes were binned into logarithmic divisions.  
Henceforth, we use the term `asset value' to refer to the average selling price of the NFT.

\section{Primer on OpenSea Platform}
\label{sec:opensea_platform}

OpenSea is the first and largest peer-to-peer marketplace for NFTs. It has attracted traders to trade assets, creators to launch their portfolios and developers to build integrated marketplaces for their applications. 

\begin{figure*}
\centering
\includegraphics[width=0.9\linewidth]{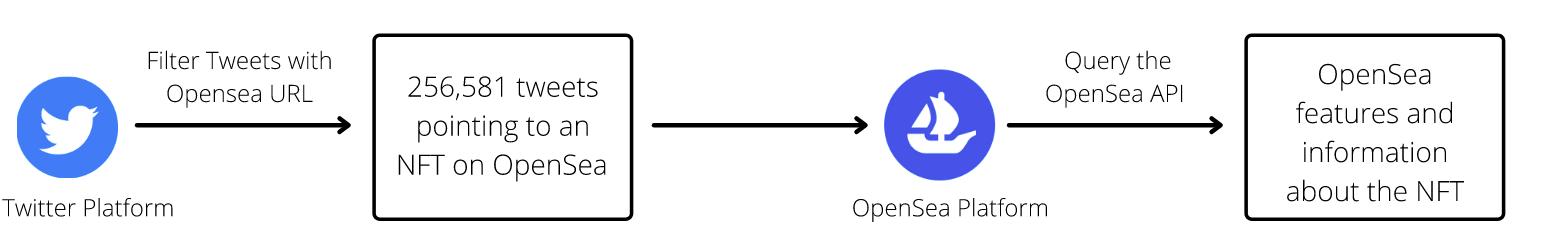}
\caption{Pipeline for the data collection process. 
}
\label{fig: data_pipeline}
\end{figure*}

The primary product on OpenSea is called the ``asset" which is a unique digital item stored as an NFT on the Blockchain. The transactions and ownership of the asset is  
programmed in a Smart Contract to store the link to the image, music or video in its metadata. Each asset is uniquely identified by its parent contract's \textit{address}, unique \textit{token id} and needs to be listed on OpenSea by the creator to be available for sale.
OpenSea permits the following transactions on an asset: (1) listing NFT at a fixed price, (2) listing NFT at first price auction or a Dutch auction, (3) or direct offers from buyers.

Once an asset is sold to a buyer, the buyer can resell it; thus, the asset keeps changing owner and price over time.
Assets on OpenSea can be further grouped into \textit{collections}. 
Collections are a group of homogeneous assets sharing common traits and properties. 
For example, the OpenSea page for one of the most popular collections, CryptoPunks (Fig.~\ref{fig: Cryptopunks}) contains similar assets with small variations. 


Most transactions on OpenSea are done on the Ethereum blockchain. Since every transaction on Ethereum requires a transaction fee, called gas-fee, accepting offers and buying assets on OpenSea are also associated with a gas-fee in addition to asset price. 

\section{Data Collection and Analysis}
\label{sec:data}
\subsection{Data Collection}
\label{sec:datacollection}
We collect data from two platforms: Twitter and OpenSea marketplace. We use the unique URL of the OpenSea asset to link these two data sources. 
Our data collection pipeline is illustrated in Fig.~\ref{fig: data_pipeline}.

\subsubsection{Twitter Data}
First, we curate 245,159 tweets from Jan 1, 2021, to Mar 30, 2021, that contain an opensea.io NFT asset link. A total of 17,155 unique users posted these tweets. Multiple tweets can reference the same OpenSea link. 
Our dataset contains 62,997 unique OpenSea assets belonging to 16,001 unique collections. 
We also collect additional information about the tweet (number of likes, number of retweets, tweet timestamp) and the source of the tweet (number of followers, number of followings, bio, account creation date). 
The additional meta-information allows us to model and understand the users posting about NFTs on Twitter.

\subsubsection{OpenSea Data}



Apart from the social media information extracted from Twitter, the OpenSea platform also provides us with many valuable features about the asset. We keep only those assets created between Jan 1, 2021, to Mar 30, 2021 (same as our tweet collection period). We used the OpenSea API\footnote{{https://docs.opensea.io/}} to extract this additional information about the asset and its collection.
Besides these features, we also crawl the associated NFT image. If the asset is a video or a GIF, we only extract and use the first frame. Overall, the dataset contains 62,997 images corresponding to unique assets.



\subsubsection{FAIR Dataset Principles}
\label{ref:fair}
The gathered data consists of publicly available information about a social network, collecting and examining which would provide significant insights into the platform's characteristics. Our dataset also conforms to the FAIR principles. In particular, the dataset is ``findable'', as it is shared publicly. This dataset is also ``accessible'', given the format used (CSV) is popular for data transfer and storage. This file format also makes the data ``interoperable'', given that most programming languages and software have libraries to process CSV files. Finally, the dataset is ``reusable'', as the included README file explains the data files in detail. The data was collected through public API endpoints of OpenSea and Twitter, adhering to their privacy policy. The data we collected was stored in a central server with restricted access and firewall protection. 

\subsection{Twitter and OpenSea Interaction Analysis}
\label{sec:twitter_opensea_linkage}
We first study the interaction between user activity on the OpenSea and Twitter platform and its influence on the asset value. We perform a temporal analysis of our dataset across both platforms. We also perform a basic correlation analysis of signals like average number of followers with asset price. 


\subsubsection{Correlation between NFT popularity across platforms} 

\begin{figure}[!b]
      \centering
\includegraphics[width=0.8\columnwidth]{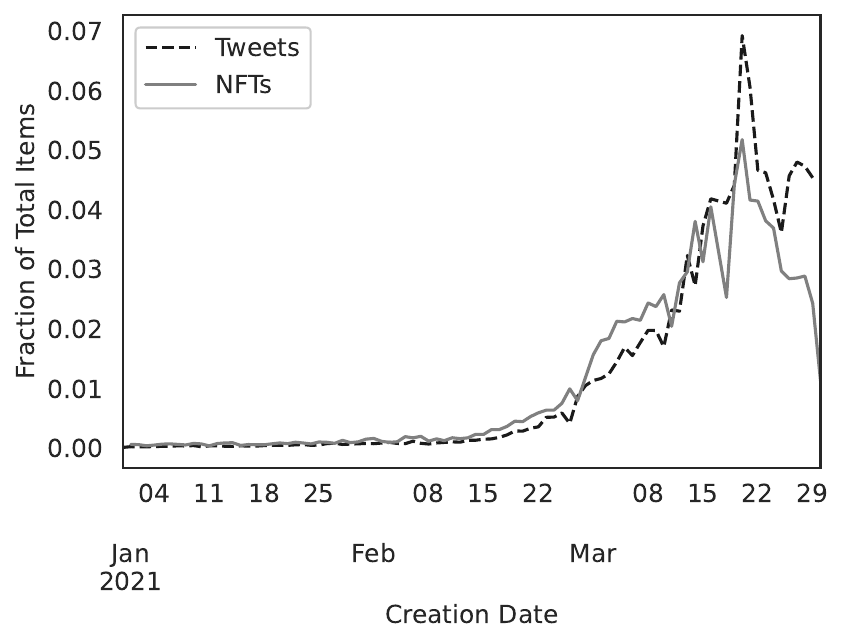}
\caption{Fraction of total tweets and NFTs created daily over the three month period from 1st Jan to 30th Mar 2021 versus asset creation dates. We observe a strong correlation between the two timeseries,
indicating that users post on Twitter soon after creating an NFT on
OpenSea.
} 
\label{fig:timeline}
    \end{figure}

We measure the NFT popularity on Twitter by aggregating features of all tweets that mention the asset. 
We have 245,159 tweets in our dataset, with the majority 89\% of them being made in March 2021.   
We plot the daily number of tweets, and the NFT asset creation dates in Fig.~\ref{fig:timeline}. The Spearman's correlation coefficient ($\rho$) for the two timeseries is 0.85 (p-value $<$ 0.001), showcasing the strong positive correlation. More than half (54.6\%) of the tweets are posted less than a day after the NFT creation.



\subsubsection{Delay between posts on Twitter and OpenSea} 

\begin{figure}
      \centering
\includegraphics[width=0.75\columnwidth]{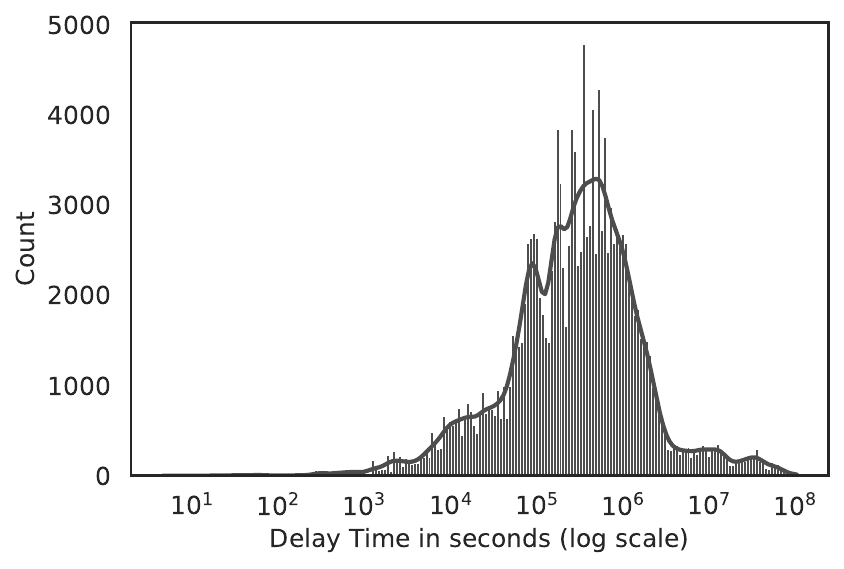}
\caption{Histogram of delay in seconds between creation of NFTs and the corresponding tweet. The graph follows an approximate log normal distribution.}
\label{fig:timelag}
    \end{figure}

Next, we analyzed the time delay between the NFT creation date on OpenSea and its mention on Twitter. 
We plot the histogram of the delay between asset creation on OpenSea and its promotion on Twitter in Fig.~\ref{fig:timelag}. The distribution of the time delay approximately follows a log normal distribution. Such distribution has been observed in other studies on online information spread~\cite{doerr_lognormal_2013} and inter-activity times~\cite{interactivity}. The inter-activity time is the duration between two consecutive tasks, like addition of followers on social media or sending of emails~\cite{interactivity}. We also analyzed the impact, the delay can have on the asset value but we found no significant correlation ($\rho$ $<$ 0.05).

\subsubsection{Twitter Username Analysis}
Next, we characterize the users on Twitter that post about these NFTs by analyzing their usernames. 
Twitter username is a strong indicator of affinity towards a cause or an organization. We computed character n-grams from usernames and manually inspected the most frequent ones. The most frequent relevant 3-gram turned out to be `nft' which was present in 7.6\% of the usernames. The other significant n-grams that came up were `crypto', `collect', and `design'. 

\begin{figure}
    \centering
\includegraphics[width=0.8\columnwidth]{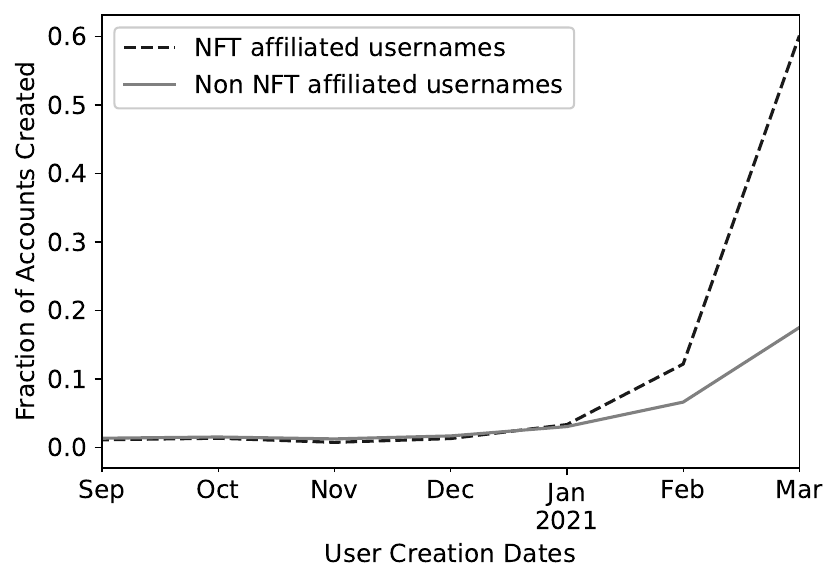}
\caption{Creation date of NFT affiliated and non NFT affiliated accounts. We see that a far higher percentage of NFT affiliated accounts were created in 2021 compared to non NFT affiliated accounts.}
 \label{fig:creation_date}
\end{figure}

Further, we partition users into two buckets: NFT affiliated (having `nft' in their username) and non NFT-affiliated, and check their account creation dates. Fig.~\ref{fig:creation_date} shows that a large proportion of NFT affiliated accounts were created in the first quarter of 2021 indicating that they were specifically created to promote and push NFT related content. Also, over 60\% of all NFT-affiliated accounts were created in March 2021, compared to only 18\% of non NFT-affiliated accounts. We also found that the mean value of assets promoted by non NFT-affiliated usernames was marginally more ($\approx$ 25\%) than those by NFT-affiliated usernames. We attribute this to the large number of low quality accounts created to specifically push NFT content. 


\subsubsection{Asset Value Analysis} 
\begin{figure}
    \centering
\includegraphics[width=0.75\columnwidth]{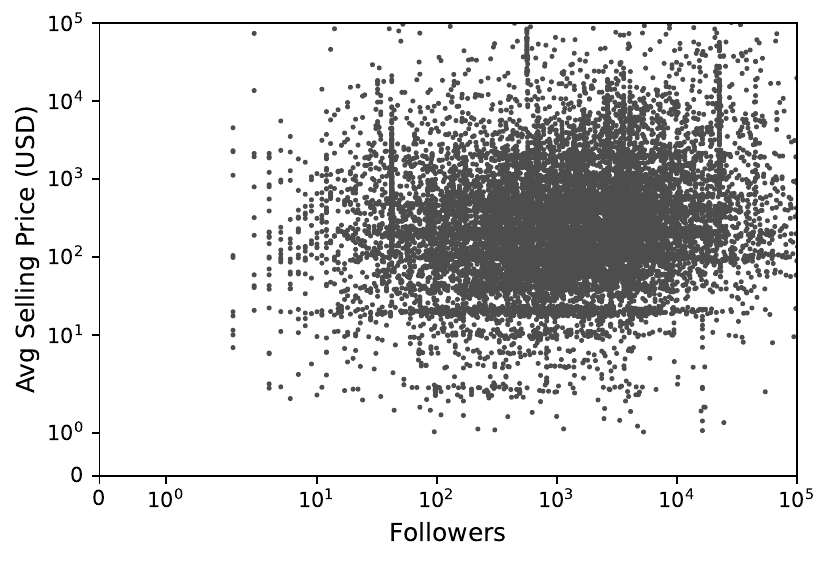}
\caption{Scatter plot for average follower count and asset value. We observe a weak positive correlation which suggests that an increase in the number of followers leads to a greater asset value.}
\label{fig:price_follow_corr}
    \end{figure}

To understand the relationship between the asset value and the popularity of the user, we plot the average number of followers (our proxy for user popularity) and the asset value of NFT in Fig \ref{fig:price_follow_corr}. Since many users can promote an NFT, we took the average follower count of the users who tweeted about each NFT. Both asset value and number of followers are highly skewed; hence we create the log-log plot. Note that we filtered out the points with asset value and follower count less than one. 

The Spearman's coefficient between the follower count and asset value on the log scale is 0.20, which indicates a weak positive correlation. 
The correlation, albeit weak, leads us to believe that social media features can help in asset value prediction. The following section discusses how we leverage features across Twitter and OpenSea platforms for accurate asset value prediction.
 

\begin{table*}
\centering
 \begin{tabular}{|l|l|p{4in}|}
\hline
\textbf{Feature} & \textbf{Type} & \textbf{Description} \\ \hline
listed count & Twitter (Account Level) & The number of accounts who have added the account to their lists \\ \hline
\hline
has nft in username & Twitter (Account Level) & True if any of the users who tweet about the NFT asset have `nft' in their screenname or username \\ \hline
n\_likes & Twitter (Tweet Level) & The number of likes on the tweet \\ \hline
n\_replies & Twitter (Tweet Level) & The number of replies on the tweet \\ \hline
n\_hastags & Twitter (Tweet Level) & The number of hashtags used with the tweet \\ \hline \hline
is presale & OpenSea & If the asset is available for pre-sale \\ \hline 
verified asset & OpenSea& If the asset is asset is verified by OpenSea\\ \hline
bid withdrawn & OpenSea & Binary label if a bid has been revoked for the asset 
 \\ \hline
bid entered & OpenSea& Number of bids placed on the asset
 \\ \hline
offer entered & OpenSea  & Number of buy offers on the asset
 \\ \hline
transfer & OpenSea  & Number of times the asset is transferred between owners
 \\ \hline
\end{tabular}
\caption{\label{table:Twitter_features}Salient features across both Twitter (account level and tweet level) and OpenSea (asset level).}
\end{table*}

\section{Features and Models for Asset Valuation}
\label{sec:Models}

We train multiple machine learning and deep learning models using a mix of Twitter, OpenSea and Image features to predict the value of an NFT asset.

\subsection{Features}
We use 77 Twitter features and 19 OpenSea features. 
The salient features from all three aspects are captured in Table~\ref{table:Twitter_features}. In the following, we discuss these features in detail.

\subsubsection{Twitter Features}
\label{sec:Twitter features}
In our dataset, 17,155 users made a total of 245,159 tweets mentioning 62,997 OpenSea NFT assets. Multiple tweets could mention the same assets; we thus aggregate the tweet level properties such as likes, replies, and retweets. Similarly, each NFT could be mentioned by multiple users; hence user account level Twitter attributes like followers counts, listed count, favorites were also aggregated across users. We used multiple aggregation functions (mean, max, min) to create a diverse feature set.  

\subsubsection{OpenSea Features}
\label{sec:OpenSea features}
We retrieve features about the NFT from the OpenSea platform. 
We use features like asset creation date, bid withdrawn and bid entered. Additionally, we engineer and gather several asset features like number of sales, number of bids and number of offers based on events like auctions, sales, offers, bids and transfers using the events endpoint of the OpenSea API. `Offers' and `Bids' both indicate an intent of purchase. Interested buyers can make offers to buy an asset with their desired amount. If their offer is accepted by the seller then the asset is transferred directly to their digital wallet. Buyers can make bids while an asset is on auction and the asset is transferred to the highest bidder on the expiration date. 

\subsection{Problem Settings: Binary vs Ordinal}

We model the NFT asset valuation problem in two ways - a binary and an ordinal multiclass classification problem. 

\subsubsection{Binary Classification}
\label{sec:binary divisions}
Our first objective is to gauge if selling an NFT can be a profitable venture. In cases when the NFT remains unsold, or is valued at a nominal amount, users are not able to cover the mandatory gas fees. In our dataset, 78\% of assets were unsold or sold for less than \$10. 
They form the loss bearing NFT class, while the rest of the assets sold for more than \$10 form the profitable NFT class.

\subsubsection{Multiclass Ordinal Classification} 
\label{sec:multiclass divisions}
Next we consider only the profit bearing class and discretize it into further subclasses.
The maximum selling NFT was found to be orders of magnitude more expensive (\$552,621) than \$10. 
We thus propose a multiclass classification, with 5 logarithmically binned classes by powers of 10. For example, Class1 includes assets selling between \$10 and \$100, Class 2 between \$100 and \$1000, and so on. Of the 14,814 total assets with sales $>$ \$10, distribution per class is as follows: Class1 (4000), Class2 (8391), Class3 (2189), Class4 (211), Class5 (23).

We observe that our classes denote asset value intervals with an intrinsic order. In terms of asset value we observe the following order, Class 1 $\prec$ Class 2 $\prec$ Class 3 $\prec$ Class 4 $\prec$ Class 5. Since the classes are not independent, treating NFT asset valuation as a nominal multiclass classification problem is not entirely accurate. We hence model the problem as an ordinal classification problem.

Formally speaking, for a $k$ class ordinal classification, we create $k-1$ binary classifiers, where the $i$-th classifier predicts the probability P(X > $i$) for every NFT asset $X$.
We then use these classifiers to compute the probability of an asset belonging to a certain class. In our case of a 5-class ordinal classification, we train 4 binary classifiers. We compute the following probabilities: 
\begin{align*}
    P(X=1)  \ &= \  1 - P(X>1),\\
    P(X=2)  \ &= \  P(X>1) - P(X>2),\\
    P(X=3)  \ &= \ P(X>2) - P(X>3),\\
    P(X=4)  \ &= \  P(X>3) - P(X>4),\\
    P(X=5)  \ &= \ P(X>4)
    \label{eq:eq1}
\end{align*} 

We assign the asset to the class with the highest probability score. We have established that ordinal classification has an intrinsic order between the classes. Due to this, standard error measures such as accuracy, precision, recall, F1-score and Mean Squared Error either ignore information (ordering of classes) or assume additional information (absolute distance between classes). We thus compute additional metrics to gauge our model performance thoroughly.
Cardoso et al.~\cite{cardoso_measuring_2011} introduced a novel ordinal classification index for measuring the performance of ordinal classification and we use it to evaluate our models. The proposed coefficient conveys how far the outcome deviates from the ideal prediction and how inconsistent the classifier is with respect to the relative order of the classes. The range of values for the ordinal classification index is [0, 1] and smaller values indicate better ordinal classification.


\subsection{Models}
We experiment with multiple machine learning as well as deep learning models. 
\subsubsection{Traditional machine learning models}
For both of the problems setups (binary as well as ordinal), we experiment with several machine learning models like logistic regression, SVM, random forests, lightGBM and XGBoost using Twitter and OpenSea features.

\begin{table*}
\centering
  \begin{tabular}{|l|l|l|l|l|}

\hline
\textbf{Feature Set Used} & \textbf{Binary Accuracy} & \textbf{Binary F1} & \textbf{Ordinal Accuracy} & \textbf{Ordinal Index} \\
\hline
\hline
\textbf{Twitter} & 83.75 & 82.00 & 67.03 & 0.3645\\ 
\hline
\textbf{OpenSea} & - & - & 63.22 &  0.3882 \\ 
\hline
\textbf{Twitter + OpenSea} & - & - & 69.33 & 0.3423 \\ 
\hline
\end{tabular}
 \caption{\label{tab:model_results} XGBoost classifier accuracy scores for different combination of features sets. Twitter features combined with OpenSea increased accuracy by 6\% showing the importance of Twitter features. The lower the ordinal index the better. For all other metrics, the higher the value, the better. Binary scores are not reported for models using OpenSea features since most OpenSea features contain information about the asset sale.
 }
\end{table*}

\subsubsection{CNNs for Image-based Predictions}
Besides Twitter and OpenSea features, we also attempt to capture the influence of the NFT image in determining the asset value of the NFT on OpenSea. Here, we predict asset value using Convolutional Neural Network (CNN)-based classifiers. 

We experiment with two CNN architectures: ResNet-101 and DenseNet-121. We feed the representative image for the NFT asset as input. We train 4 binary classifiers and apply the Softmax function on the final layer to obtain class probabilities as discussed in Section~\ref{sec:multiclass divisions}.

For each model, we use the Adam optimizer with a learning rate of 0.001. 
The images are augmented using Random Affine and Random horizontal flip to avoid model overfitting. The image pixel values are normalized and scaled to values between -1 to 1 and are grouped into batches of size 128. We use the cross-entropy loss.

\section{Asset Valuation Results}
\label{sec:results}
We tried multiple machine learning models like SVM, Logistic Regression, and Decision tree-based ensemble models like Random Forests and XGBoost (eXtreme Gradient Boosting). We use a 75\% training and 25\% test split for all the models. We performed several experiments by varying the classifier, problem setup (binary vs ordinal) and feature sets. Across all such combinations, XGBoost Classifier performed the best. For brevity purposes, we focus on results from XGBoost in this section. We use the average gain metric to compute XGBoost feature importance. It measures gains across all splits where the feature was used to assign importance score. All the results in the following sections use the XGBoost classifier and the average gain as the feature importance metric.  

\subsection{Accuracy with Different Feature Sets}
Table~\ref{tab:model_results} shows the binary/ordinal classification accuracy and the ordinal classification index using various feature set combinations.

\noindent\textbf{Tweet Features}: Using just tweet features, we obtained an accuracy of 83.75\% for the binary classification and 67.03\% for the ordinal classification problem using XGBoost. The ordinal classification index for the model is 0.36. We observe that features like listed count, number of likes and replies and presence of NFT in Twitter username are the most important.

\noindent\textbf{OpenSea Features}: 
The classifier provided an accuracy of 63.22\% for the ordinal classification problem with an ordinal index score of 0.39. Since the asset level features include signals like number of sales, using these features for binary classification leads to almost perfect accuracy values. Thus, we report only ordinal classification accuracies when using OpenSea asset features in Table~\ref{tab:model_results}. 
It is interesting to note that even Twitter features alone could predict the asset value better than the inherent asset features. This indicates that social media features highly influence the value of an NFT. 


\begin{figure}[!b]
    \centering
     \includegraphics[width=\linewidth]{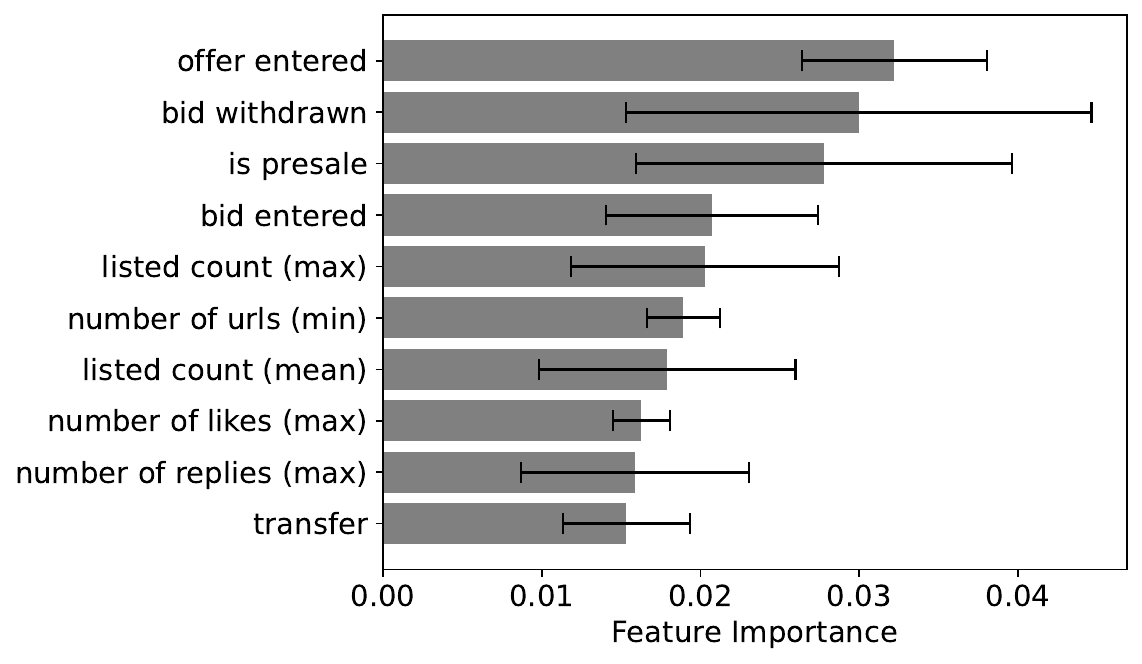}
     \caption{Feature importance scores of Twitter + OpenSea ensemble XGBoost model. We observe a mix of both OpenSea (offer entered, bid withdrawn, is presale) and Twitter (listed count, maximum listed count, maximum number of likes, replies) features indicating both the feature sets help in the prediction.
     }
     \label{feature_importance}
\end{figure}

\noindent\textbf{Twitter + OpenSea  Features}: 
\label{sec:Ensemble}
Finally, we combine Twitter features along with OpenSea features. The ensemble model of Twitter + OpenSea features performs better than either of them individually with a final accuracy of 69.33\% for the ordinal classification task, while the OpenSea features alone show an accuracy of 63.22\%. There is a significant improvement of over 6 absolute percentage points in accuracy and $\sim$0.05 points from 0.39 to 0.34 in ordinal classification index when compared to using only OpenSea features. We computed feature importances and found that both the Twitter and the OpenSea features appear in the top feature importance list as shown in Fig.~\ref{feature_importance}. 
We also observe that Twitter features like listed count, maximum listed count, maximum number of likes, replies have high importance scores. On the other hand, OpenSea features like offer entered, bids withdrawn, bid entered and is presale are important as well.

\subsection{Accuracy using Different Classifiers}
The scores for different models for the best performing Twitter + OpenSea ensemble are listed in Table~\ref{tab:different_model}. We observe that XGBoost leads to much better ordinal accuracy and the lowest ordinal index amongst all classifiers.

\begin{table}
\begin{tabular}{|l|l|l|}
\hline
\textbf{Model} & \textbf{Ordinal Accuracy} & \textbf{Ordinal Index} \\ \hline
\hline
Logistic Regression & 60.23 & 0.4049 \\ \hline
SVM & 63.90 & 0.3880 \\ \hline
Random Forest & 66.31 & 0.3459 \\ \hline
LightGBM & 68.84 & 0.3441 \\ \hline
\textbf{XGBoost} & \textbf{69.33} & \textbf{0.3423} \\ \hline
\end{tabular}
\caption{\label{tab:different_model} Ordinal accuracy and index for the final Twitter + OpenSea ensemble. XGBoost outperforms the other models. Similar trend was observed with other features sets. Lower ordinal index and higher  ordinal accuracy are better.}
\end{table}

\subsection{Accuracy with CNNs}
We obtain accuracies of 54.62\% and 52.94\% with pretrained DenseNet-121 and ResNet-101 models respectively, for the multiclass ordinal classification task, which is lower than that of the Twitter model (67.03\%) as well as OpenSea model (63.22\%). Even in terms of the ordinal classification index values, we observe that the image-based model reports values of 0.46 and 0.43 respectively for Resnet-101 and Densenet-121 architectures, which is significantly worse than the models based on Twitter features (0.36), OpenSea features (0.39) and even their ensemble (0.34). The ResNet-101 model has an accuracy of 79.01\% and the DenseNet-121 model has an accuracy of 79.44\% in the binary classification setting.

Regardless of the architecture used, the models using image information display lower accuracies and F1-scores and higher ordinal classification index values than those using Twitter and OpenSea features and their ensembles described in Table~\ref{tab:model_results} under identical training and validation settings. On adding image features to the Twitter-OpenSea ensemble, we notice no improvement in model performance. The lack of increase in accuracy along with poor results on using the image features independently reinforce our initial hypothesis that the branding and context surrounding an NFT influences the asset value more than the content itself.
\section{Conclusion}
\label{sec:conclusion} 
We have seen in the past how `digital' has triumphed over the `physical' with online advertisements, e-commerce, streaming services becoming more prevalent than their physical counterparts. Similarly 
NFTs have the potential to challenge the collectible and art market worth over \$350 Billion. In this paper, we track the growth of NFTs and show how social media reach can impact its value. We build models to predict asset value and find that Twitter features listed count and username characteristic significantly affect the value of an asset. We layout the first work to characterise and value NFT assets using social media features. Our proposed system can be used to build a profitable trading strategy by identifying overvalued and undervalued assets.   

NFT marketplace is in its nascent stage with limited data compared to well established financial markets for instruments like stocks, bonds, and options. Hence utilisation of large data-hungry deep learning models is not viable. NFTs are a fast evolving industry with numerous challenges appearing daily. For example, on 28th October, an NFT asset - CryptoPunk 9988 was sold for over \$532 Million USD.\footnote{https://www.bloomberg.com/news/articles/2021-10-29/here-s-a-532-million-nft-trade-that-wasn-t-what-it-appeared} However, the buyer and seller of the asset was the same person and the goal was to artifically inflate prices and become viral on social media. Our current system would not handle such outlier transactions and schemes. A better understanding of the security issues surrounding NFTs~\cite{security_nft_ddas} will enable us to create more robust systems. In future, we plan to build systems to automatically detect such fraudulent transactions that try to create artificial market signals. 

\bibliographystyle{splncs04}
\bibliography{references}
\end{document}